# Notes on the mathematical basis of the UK Non-Native Organism Risk Assessment Scheme[1]

Gareth Hughes
School of Biological Sciences, University of Edinburgh, Edinburgh EH9 3JG, U.K.

"*It follows from the international legal frameworks for biosecurity … that decision-making depends upon an adequate infrastructure for risk analysis.*" Outhwaite et al. (2008)

## 1. Introduction

In 2004, the UK Government's Department for Environment, Food and Rural Affairs (DEFRA) commissioned research with the objective of developing a scheme for assessing the risks posed to species, habitats and ecosystems in the UK by non-native organisms. The outcome was the UK Non-Native Organism Risk Assessment Scheme[2][3]. A comprehensive overview of the risk assessment methodology and its application is provided by Baker et al. (2008). Unfortunately, the mathematical basis of the procedure for summarising risks and uncertainties deployed in the Risk Assessment Scheme, as outlined in Baker et al. (2008, section 3.5) and described in more detail in two further papers (Holt, 2006; Holt et al., 2006), is problematic. The problems are discussed in the notes that follow.

## 2. Some notation and terminology

For clarity, we need to expand the notation used by Holt (2006) and Holt et al. (2006). We use the term *experiment* to refer generically to all the procedures used to generate a set of data, whether the data are observational or hypothetical. The common objective of the three experiments discussed here is to provide a basis for prediction of the *level of threat* presented by the import of alien species for indigenous species and habitats. In an experiment, a number of independent *components of risk* related (or at least, thought to be related) to the level of threat posed by a number of alien species are scored. The scale on which scores for risk components are coded is calibrated so that higher scores are associated with greater level of threat. Depending on the design of the experiment, the true threat status of the species involved may or may not be known.

In an experiment, the maximum number of risk components that may be scored is denoted $R$, and a particular component is identified by the subscript $i$, as $R_i$. A total of $T$ species are assessed, but not all risk components are necessarily scored for each species. A particular species is denoted by the subscript $j$, as $T_j$. For the $j$th species ($j=1…T$), the total number of risk components scored is denoted $R_j$ ($1≤R_j≤R$), where necessary. We can refer to a particular risk component for a particular species, where necessary, as $R_{ij}$ ($i=1…R_j$). The level of threat ($v$) is indicated by the score $s(v)$, which is typically coded on an *n*-point ordinal-categorical scale, using a subscript $k$ ($k$=1…$n$) to denote a particular score, as $s_k(v)$. Thus we can refer to the particular score (code) given to a particular species for a particular risk component, where necessary, as $s_{ijk}(v)$.

We denote probabilities as Pr(•). To convert from the probability Pr(•) of an event to the odds Odds(•) of an event, we note that Odds(•) = Pr(•)/(1− Pr(•)) (so that the operational precedence is clear). For the reverse process, the conversion is Pr(•) = Odds(•)/(1+ Odds(•)) (again, for clarity of operational precedence). Holt (2006) and Holt et al. (2006) use the terms *threat* and *risk* apparently interchangeably, and neither is defined. Clearly, both terms are meant to convey the impression that the import of alien species may have undesirable effects on indigenous species and habitats. Here, threat is used when the actual or predicted status of an alien species is measured on an ordinal-categorical scale. Risk is used as a synonym for probability, specifically when the event in question an undesirable one.

## 3. Experiment 1

In Experiment 1 (described on page 61 of Holt (2006)), a total of $T = 40$ species were assessed, of which 25 were definitively judged to be threats (denoted $V$) and 15 definitively non-threats (denoted $\neg V$). These judgements were made independent of the scoring of risk components and represent the true status of each species. We will assume that the 40 species have in common some characteristics (of geography, taxonomy or ecology, for example) that make it a sensible idea for us to consider them as a natural group, and note that application of the results of the analysis in future should be restricted to species that share the same group characteristics. There are no problems with Experiment 1 as described by Holt (2006). The experiment is described here, in some detail, for subsequent comparison with Experiments 2 (Holt et al., 2006) and 3 (Holt, 2006).

The data presented comprise two separate (partial) frequency distributions of scores (for threats and non-threats) for a single risk component (so $R = 1$ in this example). For each species, scores for this risk component were coded on a 5-point scale ($n$=5, $k$=1…$n$, $s_k(v)$=0…4), calibrated so that higher scores are associated with greater level of threat. Interpretations of the five points on the scale were as follows:

| $k$ | $s_k(v)$ | Predicted level of threat |
|---|---|---|
| 1 | 0 | very low |
| 2 | 1 | low |
| 3 | 2 | neutral |
| 4 | 3 | high |
| 5 | 4 | very high |

Thus, Experiment 1 is essentially designed to find a quantitative relationship between predicted level of threat and the true threat status. This relationship will then provide a basis for predicting the threat status of new species presented for risk assessment, for which the true status is unknown and any regulatory action must be taken on the basis of the predicted status. The data presented for Experiment 1 were as follows ('?' represents data not presented):

| | | True status | |
|---|---|---|---|
| $k$ | $s_k(v)$ | Non-threat ($\neg V$) | Threat ($V$) |
| 1 | 0 | ? | ? |
| 2 | 1 | ? | ? |
| 3 | 2 | ? | ? |
| 4 | 3 | ? | ? |
| 5 | 4 | 6 | 14 |
| | Sums | 15 | 25 |

### 3.1. Analysis of Experiment 1

From these data, we can calculate $\Pr(s_5(v)|V) = 14/25 = 0.56$ and $\Pr(s_5(v)|\neg V) = 6/15 = 0.40$ . We define the quantity $L_k = \Pr(s_k(v)|V)/\Pr(s_k(v)|\neg V)$ and note that $L_k$ is an *interval likelihood ratio* (as discussed by, for example, Brown and Reeves (2003) and Sonis (1999)). Here, for $k$=5, $L_k = 0.56/0.40 = 1.40$. For brevity, $L_k$ is often just referred to as a likelihood ratio. "*The likelihood ratio* [$L_k = 0.56/0.40 = 1.40$] *means that it was* 1.4 *times as likely that* [the risk component under consideration] *would be given a score of* '4' *if it was a threat*" (Holt, 2006). Just to be clear on this, the likelihood ratio $L_k = 1.40$ (for $k$=5 in this example) means that scores of $s_k(v)$=4 for the risk component under consideration are 1.4 times as likely to come from species that are threats than from species that are non-threats (see, for example, Sackett et al., 1991).

Together, the $L_k$ ($k$=1…5) would constitute a discrete *likelihood ratio function*. The data presented by Holt (2006) do not allow the calculation of the rest of the function, but we can present the above data in abbreviated classification, as follows:

| | | True status | |
|---|---|---|---|
| $k$ | $s_k(v)$ | Non-threat ($\neg V$) | Threat ($V$) |
| 1 | 0-3 | 9 | 11 |
| 2 | 4 | 6 | 14 |
| | Sums | 15 | 25 |

In this abbreviated classification, $\Pr(s_1(v)|V) = 11/25 = 0.44$ and $\Pr(s_1(v)|\neg V) = 9/15 = 0.60$. $\Pr(s_2(v)|V) = 14/25 = 0.56$ and $\Pr(s_2(v)|\neg V) = 6/15 = 0.40$. We can now calculate the two likelihood ratios:

$L_1 = \Pr(s_1(v)|V)/\Pr(s_1(v)|\neg V) = 0.44/0.60 = 0.73$
$L_2 = \Pr(s_2(v)|V)/\Pr(s_2(v)|\neg V) = 0.56/0.40 = 1.40$

The significance of these likelihood ratios is that they allow us to update the prior probability of threat (or non-threat) posed by a new alien species. Thus we can obtain the posterior probability of threat (or non-threat), having taken into account the evidence based on assessment of (in this case) a single risk component. This updating is made via Bayes' theorem. We begin with two prior probabilities, $\Pr(V)$ and $\Pr(\neg V)$ (such that $\Pr(V) + \Pr(\neg V) = 1$). Following assessment of the evidence, we have four posterior probabilities: $\Pr(V|s_2(v))$ and $\Pr(\neg V|s_2(v))$ (such that $\Pr(V|s_2(v)) + \Pr(\neg V|s_2(v)) = 1$), and $\Pr(\neg V|s_1(v))$ and $\Pr(V|s_1(v))$ (such that $\Pr(\neg V|s_1(v)) + \Pr(V|s_1(v)) = 1$).

### 3.2. Results of Experiment 1

Using Bayes' theorem, we can obtain:

$\text{Odds}(V|s_2(v)) = \text{Odds}(V) \times L_2$
$\text{Odds}(\neg V|s_2(v)) = \text{Odds}(\neg V) \times L_2^{-1}$
$\text{Odds}(\neg V|s_1(v)) = \text{Odds}(\neg V) \times L_1^{-1}$
$\text{Odds}(V|s_1(v)) = \text{Odds}(V) \times L_1$

A value of $\Pr(V)$ is not supplied in the description of Experiment 1, but we can follow Holt (2006) and take $\Pr(V) = 0.5$ (and so $\Pr(\neg V) = 0.5$), for the purpose of illustration. Prior probabilities are often based on historical data, but Holt (2006) justifies the adoption of $\Pr(V) = \Pr(\neg V) = 0.5$ by noting that results obtained on this basis are considered to be relative, rather than absolute, measures of risk. Then $\text{Odds}(V) = \text{Odds}(\neg V) = 1$ and:

$\text{Odds}(V|s_2(v)) = 1.400$
$\text{Odds}(\neg V|s_2(v)) = 0.714$
$\text{Odds}(\neg V|s_1(v)) = 1.364$
$\text{Odds}(V|s_1(v)) = 0.733$

and then:

$\Pr(V|s_2(v)) = 0.583$
$\Pr(\neg V|s_2(v)) = 0.417$
$\Pr(\neg V|s_1(v)) = 0.577$
$\Pr(V|s_1(v)) = 0.423$

An alternative view would be possible if it were justifiable to regard the sample of species on which the analysis is based as representative of the population of species to which the results of the analysis will be applied. In that case we could take the relative frequencies of threat ($V$) and non-threat ($\neg V$) as estimates of the prior probabilities. Then $\Pr(V) = 25/40 = 0.625$ and $\Pr(\neg V) = 15/40 = 0.375$. Using Bayes' theorem as above leads to the posterior probabilities:

$\Pr(V|s_2(v)) = 0.70$
$\Pr(\neg V|s_2(v)) = 0.30$
$\Pr(\neg V|s_1(v)) = 0.45$
$\Pr(V|s_1(v)) = 0.55$

### 3.3. Discussion of Experiment 1

Suppose that a new alien species is presented for risk assessment. It is not part of the experiment, so we do not know its true threat status, but we have the score $s_k(v)$ for the single risk component, and appropriate estimates of the prior probabilities $\Pr(V)$ and $\Pr(\neg V)$. The posterior probabilities $\Pr(V|s_k(v))$ and $\Pr(\neg V|s_k(v))$ ($k$=1,2) are then evidence-based revisions of prior probabilities. Finally, of course, for this kind of risk assessment scheme to become operational, we also need to have specified in advance what actions will follow once the risks (as posterior probabilities of threat status, given the evidence) have been identified by the analysis.

Suppose that the analysis of data for a second risk component were available (so $R$=2). We need a bit more of the notation now: the two risk components are denoted $R_i$ ($i$=1,2) and the $k$th likelihood ratio resulting from analysis of $R_i$ is $L_{ik}$ ($k$=1…$n$). The risk component scores for a new alien species presented for risk assessment are $s_{1k}(v)$ for $R_1$ and $s_{2k}(v)$ for $R_2$ ($k$=1…$n$). For the set of scores $\{s_{ik}(v)\}$, $\mathrm{Odds}(V|\{s_{ik}(v)\})$ = $\mathrm{Odds}(V) \times L_{1k} \times L_{2k}$ (see, for example, Go (1998)). In general, for a total of $R$ risk components:

$$\mathrm{Odds}\big(V\big|\{s_{ik}(v)\}\big) = \mathrm{Odds}(V) \times \prod_{i=1}^{R} L_{ik} \qquad\qquad \text{(Equation 1)}$$

in which the subscript $k$ of each individual score $s_{ik}(v)$ ($i$=1…$R$) (i.e., one score for each risk component analysed) tells us which one of the $k$=1…$n$ likelihood ratios $L_{ik}$ to include as one of the $R$ terms in the product on the right-hand side of the equation (i.e., one likelihood ratio for each score).

## 4. Experiment 2

Experiment 2 is described in Holt et al. (2006; Methodology, page 169; Results, Table 3). Data for a total of $T$ = 6 species are presented (actually, data for 256 species were analysed but only those presented in Table 3 of Holt et al. (2006) are considered here). For each species, $R$ = 7 risk components were scored. Scores were initially coded on a 3-point ordinal-categorical scale ($n$=3, $k$=1…$n$, $s_k(v)$=1,2,3) which is calibrated so that higher scores are indicative of greater level of threat.

If we compare Experiment 2 with Experiment 1, we see that in Experiment 2 the distribution of scores for a single risk component $R_i$ is given in each column of the Table 1. However, this is a distribution over all $T_j$ species. Unlike Experiment 1, there is no definitive judgement of the true threat ($V$) or non-threat ($\neg V$) status of the species in Experiment 2: all are 'potential quarantine pests'. We cannot, therefore, proceed by calculating likelihood ratios and using these to update prior probabilities to posterior probabilities, as in Experiment 1.

**Table 1.** Scores for risk components in Experiment 2.

| Species | Risk component $R_i$ ($i$=1…7) | | | | | | | Total | Arithmetic |
|---|---|---|---|---|---|---|---|---|---|
| $T_j$ ($j$=1…6) | $R_1$ | $R_2$ | $R_3$ | $R_4$ | $R_5$ | $R_6$ | $R_7$ | score | mean |
| $T_1$ | 2 | 3 | 2 | 1 | 1 | 1 | 1 | 11 | 1.57 |
| $T_2$ | 1 | 1 | 2 | 1 | 3 | 2 | 2 | 12 | 1.71 |
| $T_3$ | 2 | 2 | 2 | 1 | 3 | 1 | 1 | 12 | 1.71 |
| $T_4$ | 2 | 3 | 3 | 1 | 1 | 1 | 1 | 12 | 1.71 |
| $T_5$ | 2 | 2 | 3 | 1 | 3 | 1 | 1 | 13 | 1.86 |
| $T_6$ | 2 | 3 | 3 | 2 | 1 | 2 | 2 | 15 | 2.14 |

The design of Experiment 2 means that the basic data available for analysis of the level of threat are the frequency distributions of risk component scores for species. These scores must be combined in some way so as to provide an overall value that constitutes the predicted level of threat for use as a basis for risk assessment. In Table 1, the total score (which in this case must lie between 7 and 21) for a species $T_j$ is $\sum_{i=1}^{R} s_{ijk}(v)$. As Holt et al. (2006) point out, either this total score, or the arithmetic mean score $\left.\sum_{i=1}^{R} s_{ijk}(v) \middle/ R\right.$, is often used to provide a basis for risk assessment. The premise is simple: a higher total or arithmetic mean score taken over the $R$ risk components assessed is indicative of greater level of threat. When the number of risk components assessed is the same for all species (as in the Table 1), the arithmetic mean score is only preferable to the total score as an overall value for the predicted level of threat because it is calibrated on the same scale as the individual risk components. When the number of risk components assessed varies between species, the arithmetic mean score is preferable because it allows easier comparisons of the overall predicted level of threat between species.

Note that in using either the total score or the arithmetic mean score we have lost some information. Specifically, species $T_2$, $T_3$ and $T_4$ all have the same total score and mean score, but the distribution of scores for species $T_4$ is not the same as the distribution of scores for species $T_2$ and $T_3$.

### 4.1. Analysis of Experiment 2

The approach of Holt et al. (2006) to the analysis of Experiment 2 depends on the device of assigning (subjective) numerical odds of threat status to the 3-point ordinal-categorical scale of scores ($s_k(v)$=1,2,3) for risk components. Holt et al. (2006) write:

- "*The assigning of scores in a risk assessment is essentially a relative process in that it is often possible to predict if the risk posed by one organism is greater or less than another, and so ordinal scoring is a reasonable way to express this understanding. If odds rather than scores are assigned directly, then the odds need only express an ordinal scale of risk in the same way as scores.*"

- "*Assigning odds instead of scores allows a more rigorous probabilistic treatment of the data…*"

In order to assign numerical odds of threat status to designated points on the ordinal-categorical scale of scores, each score for risk component $R_i$ for species $T_j$ is regarded as an item of *evidence*, denoted here $s_{ijk}(v)$ ($k$=1,2,3) Assigning odds, according to Holt et al. (2006), amounts to assessing how likely is this evidence given that the species in question is a threat. In conditional odds notation, this is Odds($s_{ijk}(v)|V$).

This reasoning is problematic. We are not actually given that any species is a threat (or non-threat) in this experiment. The introduction of conditionality implies that we have some *additional information* about an event that might influence our view of its odds (or probability). However, Experiment 2 only provides information on scores for risk components. In the present context, whether the scores for risk components take the form of values on a 3-point ordinal-categorical scale or values regarded as numerical odds, neither one provides any additional information about the other. We can replace the risk component scores by (subjective) numerical odds if we want, but in so doing we do not increase the information content of our data. Note that odds are assigned only at the three points for which scores are given, so the numerical odds are regarded as being on an ordinal scale. Essentially, the outcome of the exercise is the replacement of one ordinal scale (of scores) with another (of odds).

A species-specific constant $M_j$ ($M_j > 1$) is used in calculating numerical odds of threat status[4]. Values of $M_j$ are not given in the description of Experiment 2, but it is straightforward to reverse engineer them from the data in Table 3 of Holt et al. (2006). However, Holt et al. (2006) do not offer an explanation for the differences between the values of $M_j$ for the different species. $\text{Odds}_{ijk}(v)$ ($k=1\ldots n$, $n=3$ in this case) is the odds of threat status applicable for risk component $R_i$ ($i=1\ldots 7$ here) for species $T_j$ ($j=1\ldots 6$ here). Odds are assigned as follows[5]: when $k = 1$, $\text{Odds}_{ijk}(v) = 1/M_j$; when $k=2$, $\text{Odds}_{ijk}(v) = 1$; when $k = 3$, $\text{Odds}_{ijk}(v) = M_j$. Table 1 of scores for risk component then becomes Table 2 of numerical odds for risk components.

**Table 2.** Holt's numerical odds for risk components in Experiment 2.

| Species | Risk component $R_i$ ($i=1\ldots 7$) | | | | | | | $\prod_{i=1}^{R} \text{Odds}_{ijk}(v)$ | 'probability' |
|---|---|---|---|---|---|---|---|---|---|
| $T_j$ ($j=1\ldots 6$) | $M_j$ | $R_1$ | $R_2$ | $R_3$ | $R_4$ | $R_5$ | $R_6$ | $R_7$ | |
| $T_1$ | 2.1624 | 1 | $M_j$ | 1 | $M_j^{-1}$ | $M_j^{-1}$ | $M_j^{-1}$ | $M_j^{-1}$ | 0.0989 | 0.090 |
| $T_2$ | 2.3805 | $M_j^{-1}$ | $M_j^{-1}$ | 1 | $M_j^{-1}$ | $M_j$ | 1 | 1 | 0.1765 | 0.150 |
| $T_3$ | 2.3805 | 1 | 1 | 1 | $M_j^{-1}$ | $M_j$ | $M_j^{-1}$ | $M_j^{-1}$ | 0.1765 | 0.150 |
| $T_4$ | 2.3805 | 1 | $M_j$ | $M_j$ | $M_j^{-1}$ | $M_j^{-1}$ | $M_j^{-1}$ | $M_j^{-1}$ | 0.1765 | 0.150 |
| $T_5$ | 2.1250 | 1 | 1 | $M_j$ | $M_j^{-1}$ | $M_j$ | $M_j^{-1}$ | $M_j^{-1}$ | 0.4706 | 0.320 |
| $T_6$ | 2.5714 | 1 | $M_j$ | $M_j$ | 1 | $M_j^{-1}$ | 1 | 1 | 2.5714 | 0.720 |

We now consider steps 1, 2 and 3 of the procedure for deriving "*a probability that the organism poses a risk, given the set of evidence.*" (see page 168 and Appendix 1 of Holt et al. (2006)).

- Step 1. For each risk component, estimate the odds that the score $s_{ijk}(v)$ occurs, given that the species is a threat.
    - Problem: actually, these odds cannot be conditional on threat status (as explained above).
- Step 2. The product of the odds over all risk components gives the combined odds that this set of scores $\{s_{ijk}(v)\}$ will occur, given that the species is a threat.
    - Problem: same problem as in Step 1.
    - Problem: unfortunately, multiplying odds together in this way does not produce a 'combined odds' that is interpretable in a way that is consistent

---

[4] The term *odds multiplier* is used by Holt et al. (2006) for $M_j$, but this terminology is not a good idea as the term is used in a different way by bookmakers in the UK when settling multiple bets, and this usage will make an appearance in section 4.2.
[5] Alternatively, note that $\log(\text{Odds}_{ijk}(v)) = -\log(M_j)$, 0, $+\log(M_j)$ for $k = 1, 2, 3$, respectively.

with the laws of probability. It is not clear how to interpret the quantity $\prod_{i=1}^{R} \text{Odds}_{ijk}(v)$.

- Reverse the conditionality.
    - Problem: same problem as in Step 1.
    - Actually, it is a good thing that there in no need to reverse the conditionality, because Equation 2 of Appendix 1 is incorrect (as a statement of Bayes' theorem). The correct version is Equation 1 from section 3.3, above.
    - Another bonus is that since we do not have to reverse the conditionality, we do not need to be concerned with making an assumption about the prior odds. No estimate of prior odds is needed to calculate $\text{Pr}(\{s_{ijk}(v)\})$.
- Step 3. Divide the combined odds by itself plus one. This converts the combined odds to a probability that the organism is a threat, given the set of scores $\{s_{ijk}(v)\}$.
    - Problem: since the odds could not be conditional in the first place, this conditional probability cannot be derived from the data in Experiment 2.
    - Problem: since it is not clear how to interpret the quantity $\prod_{i=1}^{R} \text{Odds}_{ijk}(v)$ (see Step 2), the resulting 'probability' $\left. \prod_{i=1}^{R} \text{Odds}_{ijk}(v) \middle/ \left[1 + \prod_{i=1}^{R} \text{Odds}_{ijk}(v)\right] \right.$ also has no clear interpretation in terms of the laws of probability.
    - These problems notwithstanding, Table 2 above shows the derivation of 'probability' as given in Table 3 of Holt et al. (2006).

Note that even if we just accept this 'probability' as some sort of summary of the numerical odds of threat status for a species, we still lose some information. Specifically, species $T_2$, $T_3$ and $T_4$ all have the same 'probability', but the distribution of numerical odds for species $T_4$ is not the same as the distribution for species $T_2$ and $T_3$.

### 4.2. How to combine odds

The easiest way to combine odds is to convert to probabilities, combine the probabilities as appropriate using the laws of probability, then convert back to odds at the end if required. Notwithstanding, if you really want to know how to combine odds, start by reading two articles by Fletcher (1994, 2004).

- We have values denoted $\text{Odds}_{ijk}(v)$ corresponding to scores $s_{ijk}(v)$ ($k$=1,2,3). We want to find a combined value for the corresponding probabilities $\text{Pr}_{ijk}(v)$, over $R$ risk components ($i$=1…$R$), for the $j$th species $T_j$. This combined value is denoted $\text{Pr}(\{s_{ijk}(v)\})$.

- Step 1: calculate $\text{Pr}_{ijk}(v) = \dfrac{\text{Odds}_{ijk}(v)}{1 + \text{Odds}_{ijk}(v)}$ ($i$=1…$R$), for the $j$th species $T_j$.

- Step 2: calculate $\text{Pr}(\{s_{ijk}(v)\}) = \prod_{i=1}^{R} \text{Pr}_{ijk}(v)$. Thus we have calculated the probability of the observed set of scores $\{s_{ijk}(v)\}$ for the $j$th species $T_j$. Of course, this is a subjective probability reflecting the subjective numerical odds of threat status assigned to the 3-point ordinal-categorical scale of scores for risk components, based on the species-specific constant $M_j$.

- Alternatively, write $\text{Odds}_{ijk}(v)$ as the bookmakers' odds $\dfrac{1}{\text{Odds}_{ijk}(v)}$ $to$ 1 (actually, bookmakers odds are usually written as 'integer $to$ integer', so, for example, if $\text{Odds}_{ijk}(v) = \dfrac{1}{2.1624}$, this is approximately '13 $to$ 6')[6].

- Calculate the corresponding *decimal odds* $\dfrac{1}{\text{Odds}_{ijk}(v)} + 1$.

- Calculate the *odds multiplier* $\displaystyle\prod_{i=1}^{R}\left[\dfrac{1}{\text{Odds}_{ijk}(v)} + 1\right]$. The odds multiplier is used by bookmakers in the UK in the calculations for settling multiple bets. At this stage we could note that $\displaystyle\prod_{i=1}^{R}\left[\dfrac{1}{\text{Odds}_{ijk}(v)} + 1\right] = \dfrac{1}{\Pr(\{s_{ijk}(v)\})}$, so we have the desired probability as in Step 2 above.

- Otherwise, write the overall bookmakers' odds as $\displaystyle\prod_{i=1}^{R}\left[\dfrac{1}{\text{Odds}_{ijk}(v)} + 1\right] - 1$ $to$ 1.

- Then $\text{Odds}(\{s_{ijk}(v)\}) = \dfrac{1}{\displaystyle\prod_{i=1}^{R}\left[\dfrac{1}{\text{Odds}_{ijk}(v)} + 1\right] - 1}$.

- Finally, $\Pr(\{s_{ijk}(v)\}) = \dfrac{\text{Odds}(\{s_{ijk}(v)\})}{1 + \text{Odds}(\{s_{ijk}(v)\})}$.

### 4.3. Results of Experiment 2

The probabilities $\Pr(\{s_{ijk}(v)\})$ are shown in Table 3. These are calculated in accordance with the laws of probability, although of course the actual values are dependent on the choice of species-specific constant $M_j$. Actually, it makes some sense to calculate the geometric mean probability $\sqrt[R]{\displaystyle\prod_{i=1}^{R}\Pr_{ijk}(v)}$. This will provide us with values that we can relate to the original probability scale for the $\Pr_{ijk}(v)$ (we note that the $\Pr(\{s_{ijk}(v)\})$ values could be very small if there were a large number of risk components). This also takes care of the problem that would arise when making comparisons between values of $\Pr(\{s_{ijk}(v)\})$ if different numbers of risk components were scored for different species.

---

[6] This refers to the way that bookmakers in the UK write odds.

**Table 3.** The probabilities $\Pr(\{s_{ijk}(v)\})$ for species in Experiment 2.

| Species $T_j$ ($j$=1…6) | $M_j$ | $\Pr(\{s_{ijk}(v)\})$ | $\sqrt[R]{\prod_{i=1}^{R}\Pr_{ijk}(v)}$ | $\sqrt[R]{\prod_{i=1}^{R}s_{ijk}(v)}$ |
|---|---|---|---|---|
| $T_1$ | 2.1624 | 0.0017 | 0.4024 | 1.4262 |
| $T_2$ | 2.3805 | 0.0023 | 0.4193 | 1.5746 |
| $T_3$ | 2.3805 | 0.0023 | 0.4193 | 1.5746 |
| $T_4$ | 2.3805 | 0.0019 | 0.4085 | 1.5112 |
| $T_5$ | 2.1250 | 0.0038 | 0.4509 | 1.6685 |
| $T_6$ | 2.5714 | 0.0091 | 0.5108 | 2.0339 |

What have we achieved? The analysis is now mathematically correct, which is a good thing, but the manipulation of the original data in the form of scores to provide numerical odds, and subsequent calculation of 'probability' achieves nothing. For the data presented in Experiment 2, we would probably be better off just calculating the geometric mean score for each species $\sqrt[R]{\prod_{i=1}^{R}s_{ijk}(v)}$ (values are given in Table 3 above). Compared to the 'probability' calculated by Holt et al. (2006), the geometric mean score for each species:

- is mathematically correct;
- is simpler to calculate;
- is more transparent (no subjective numerical odds based on an unexplained species-specific constant are required);
- is interpretable directly on the original ordinal-categorical scale of scores;
- preserves more information from the original data. Note that the $\sqrt[R]{\prod_{i=1}^{R}s_{ijk}(v)}$ (also the $\Pr(\{s_{ijk}(v)\})$ and $\sqrt[R]{\prod_{i=1}^{R}\Pr_{ijk}(v)}$) capture a difference between species $T_4$ and species $T_2$ and $T_3$. Recall that species $T_2$, $T_3$ and $T_4$ all have the same total score, arithmetic mean score and 'probability' as calculated by Holt et al. (2006), but the distribution of scores for species $T_4$ is not the same as the distribution of scores for species $T_2$ and $T_3$.

## 5. Experiment 3

Experiment 3 is described in Holt (2006; Methodology, pages 59-60; Results, Figure 1 and Tables 2 and 3). The design is rather complicated. Data for a total of $T = 4$ species are presented (see Figure 1 and Table 3 of Holt (2006)). For each species, a number of risk components were scored, but not all risk components were scored for each species. A particular species is denoted by the subscript $j$, as $T_j$. For the $j$th species ($j$=1…4), the total number of risk components scored is denoted $R_j$. Here, $R_1$=47, $R_2$=45, $R_3$=43 and $R_4$=34[7]. Where necessary, we can refer to a particular risk component for a particular species as $R_{ij}$ ($i$=1…$R_j$), but the risk components are not identified individually in the data set. The level of threat ($v$) is indicated by the score $s(v)$, which is coded on a 5-point ordinal-categorical scale, using a subscript $k$ ($k=1\ldots n$, $n=5$) to denote a particular score, as $s_k(v)$ ($s_k(v)=0\ldots4$). The scale is calibrated so that higher scores are indicative of greater level of threat. Thus we can refer to the particular score given to a particular species for a particular risk component, where necessary, as $s_{ijk}(v)$. The data presented comprise frequency distributions of risk component scores for each species.

---

[7] Values read from Fig. 1 of Holt (2006).

Of the 4 species, one was definitively judged to be *low threat* (denoted $V_1$), one was definitively *medium threat* (denoted $V_2$) and two were definitively *high threat* (denoted here $V_3$). These judgements were made by assessors, independent of the scoring of risk components, and represent the true status of each species.

### 5.1. Analysis of Experiment 3

Holt's (2006) approach to the analysis of Experiment 3 essentially ignores the data on the true threat status of each species and proceeds along the lines of the analysis of Experiment 2 (where such data were unavailable). Here, we will first examine Holt's (2006) approach, then look at an approach that incorporates data on true threat status.

Holt (2006) first assigns a probability $\Pr_k(v)$ to each point on the 5-point ordinal-categorical scale of scores, such that $\Pr_k(v) = 0.5 + [s_k(v) - s_3(v)]\,c$. The value of $c$, an empirical *conversion parameter*, was taken to be $c = 0.017$ (a constant), and $s_3(v)$ is the point on the scale of risk component scores at which the predicted level of threat is neutral. Then $\mathrm{Odds}_k(v) = \dfrac{\Pr_k(v)}{1 - \Pr_k(v)}$ and we have:

| $k$ | $s_k(v)$ | Predicted level of threat | $\Pr_k(v)$ | $\mathrm{Odds}_k(v)$ |
|---|---|---|---|---|
| 1 | 0 | very low | 0.466 | 0.873 |
| 2 | 1 | low | 0.483 | 0.934 |
| 3 | 2 | neutral | 0.500 | 1.000 |
| 4 | 3 | high | 0.517 | 1.070 |
| 5 | 4 | very high | 0.534 | 1.146 |

Holt (2006) describes a procedure for deriving "*the conditional probability that a species is a threat given the set of scores obtained.*"

- Holt (2006) regards the probability $\Pr_{ijk}(v)$ as a conditional probability $\Pr\big(s_{ijk}(v)\big|V\big)$, i.e., the probability of score $s_{ijk}(v)$ for a particular risk component for a particular species, given that the species in question is a threat (denoted here $V$).
    - Problem: actually, these odds cannot be conditional. Conditionality can't just be assigned to a probability according to how you regard it; conditionality represents additional information (see section 4.1). In fact, we do have this additional information here, in the form of the judgements of threat status made by assessors. But those are not being called on by Holt (2006) in this analysis.
- Holt (2006) then calculates the conditional probability $\Pr\big(s_{ijk}(v)\big|\neg V\big)$ of score $s_{ijk}(v)$ for a particular risk component for a particular species given that the species in question is *not* a threat (denoted here $\neg V$) by assuming that $\Pr\left(s_{ijk}(v)\big|\neg V\right) = 1 - \Pr\left(s_{ijk}(v)\big|V\right)$.

- Problem: there is no basis in the laws of probability for this assumption. The notations '$|V$' and '$|\neg V$' refer us to two different probability distributions (see the analysis of Experiment 1 (section 3.1) for a numerical example). Think of it like this: you can't regard the same score as being conditional on the species being a threat, then as being conditional on the species being a non-threat. The species is either one or the other; you can't have it both ways.
- Problem: here Holt (2006) classifies species as either a *threat* or a *non-threat*, when actually there are 3 categories of threat status (low, medium, high; see Table 3 of Holt (2006)).

- For each risk component, Holt (2006, Equation 2) calculates interval likelihood ratios $\Pr\left(s_{ijk}(v)\big|V\right)/\Pr\left(s_{ijk}(v)\big|\neg V\right)$ ($k=1\ldots5$) and says that these can be thought of as the conditional odds that a risk component will have a particular score, given that the species in question is a threat.
  - Problem: $\Pr\left(s_{ijk}(v)\big|V\right)/\Pr\left(s_{ijk}(v)\big|\neg V\right)$ is indeed the appropriate formula for calculation of interval likelihood ratios (see section 3.1). However, these can be thought of as conditional odds only because of the way that the component conditional probabilities are (wrongly) defined here. There are two errors (detailed above): first, conditionality cannot just be assigned as described by Holt (2006); second, calculation of $\Pr\left(s_{ijk}(v)\big|\neg V\right)$ by Holt (2006) is based on an invalid assumption.

- Next Holt (2006) finds, separately for each species $T_j$, the product of the $R_j$ values of $\Pr\left(s_{ijk}(v)\big|V\right)/\Pr\left(s_{ijk}(v)\big|\neg V\right)$ (each risk component contributes one value, depending on the score $s_{ijk}$).
  - Problem: in Holt's (2006) Equation 3, the product is taken over $i=1\ldots n$ (i.e., the range of scores) rather than (as it should be) $i=1\ldots R_j$ (i.e., the range of risk components for each species).
  - Problem: Holt (2006) claims that this product is the combined conditional odds Odds($\{s_{ijk}(v)\}|V$), but as in the analysis of Experiment 2 (see section 4.1), multiplying odds together in this way does not produce a 'combined odds' that is interpretable in a way that is consistent with the laws of probability.

- Holt (2006) needs to reverse the conditionality to obtain Odds($V|\{s_{ijk}(v)\}$).
  - Problem: as discussed above, the odds as described by Holt (2006) cannot be conditional. As with Experiment 2, it is a good thing that there in no need to reverse the conditionality, because here Equation 4 in Holt (2006) is incorrect (as a statement of Bayes' theorem). The correct version is Equation 1 from section 3.3.
  - And again as in Experiment 2, it is a bonus that since we do not have to reverse the conditionality, since then we do not need to be concerned with making an assumption about the prior odds. No estimate of prior odds is needed to calculate $\Pr\left(\{s_{ijk}(v)\}\right)$.

- Holt (2006) converts the combined conditional odds Odds($V|\{s_{ijk}(v)\}$) to the corresponding conditional probability that the organism is a threat, given the set of scores obtained.

- Problem: since (at least in the way the data in Experiment 3 were analysed by Holt (2006)) the odds could not be conditional in the first place, this conditional probability cannot be derived as shown by Holt (2006).

- Problem: since it is not clear how to interpret the quantity $\prod_{i=1}^{R_j} \text{Odds}_{ijk}(v)$, the resulting 'probability' $\prod_{i=1}^{R_j} \text{Odds}_{ijk}(v) \Big/ \left[ 1 + \prod_{i=1}^{R_j} \text{Odds}_{ijk}(v) \right]$ also has no clear interpretation.

### 5.2. Results of Experiment 3

All these problems notwithstanding, Table 4 below shows the 'probability' for each species, as given in Table 2 of Holt (2006). Holt (2006) does add the caveat that these are not true probabilities, but that they provide a relative measure of risk in the range $0 - 1$. However, given the problems in the analysis leading to Holt's (2006) 'probability', it is quite difficult to see what value this measure can have as a basis for risk assessment. Also in Table 4, for completeness, is the probability $\Pr(\{s_{ijk}(v)\}) = \prod_{i=1}^{R_j} \Pr_{ijk}(v)$, i.e., the combined probability of the set of scores $s_{ijk}(v)$ calculated according to the laws of probability. Finally, the geometric mean probability $\sqrt[R_j]{\prod_{i=1}^{R_j} \Pr_{ijk}(v)}$ provides us with values that we can relate to the probability scale that was attached to the original ordinal-categorical scale of scores (shown in Table 3 above).

**Table 4.** The probabilities $\Pr(\{s_{ijk}(v)\})$ for species in Experiment 3.

| Species $T_j$ ($j$=1…4) | $R_j$ | 'probability' | $\Pr(\{s_{ijk}(v)\})$ | $\sqrt[R_j]{\prod_{i=1}^{R_j} \Pr_{ijk}(v)}$ |
|---|---|---|---|---|
| $T_1$ | 47 | 0.855 | $1.640 \times 10^{-14}$ | 0.509 |
| $T_2$ | 45 | 0.892 | $7.737 \times 10^{-14}$ | 0.511 |
| $T_3$ | 43 | 0.584 | $1.282 \times 10^{-13}$ | 0.501 |
| $T_4$ | 34 | 0.306 | $3.705 \times 10^{-11}$ | 0.493 |

We have still have not achieved anything that is demonstrably better than just calculating the geometric mean score for each species (as outlined in section 4.3). Unfortunately, if the adopted scale of scores includes '0' (as in Experiment 3) (and there are zeros in the data) then the geometric mean score will also be zero. However, this difficulty could easily be overcome by adjusting the adopted scale of scores from 0,1,2,3,4 to 1,2,3,4,5.

### 5.3. Conditional probabilities in Experiment 3

Since there are data on the true threat status of the species (in Table 3 of Holt (2006)) as well as the predicted status, we can calculate conditional probabilities. The calculation is given here *only* for the purpose of illustration. It is meant to be applied to situations in which there is a representative sample of species, such that the overall frequency distribution of species among the categories of true threat status reflects the corresponding frequency distribution in the population of interest (Holt et al. (2006), where data for 256 potential quarantine pests were analysed, may be such a case). Here we will assume that all risk components carry equal weight. For Experiment 3, frequencies for risk scores are given in Table 5. In Table 6, these frequencies have been converted to proportions that are taken as probabilities for risk scores in Experiment 3.

**Table 5.** Frequencies for risk scores in Experiment 3.

| Predicted level of threat | True threat status | | | |
|---|---|---|---|---|
| | Low ($V_1$) | Medium ($V_2$) | High ($V_3$) | |
| Score | (species $T_4$) | (species $T_3$) | (species $T_1$, $T_2$) | Row sum |
| 0  (very low) | 12 | 7 | 7 | 26 |
| 1  (low) | 3 | 9 | 9 | 21 |
| 2  (neutral) | 8 | 10 | 26 | 44 |
| 3  (high) | 7 | 6 | 20 | 33 |
| 4  (very high) | 4 | 11 | 30 | 45 |
| Column sum | 34 | 43 | 92 | 169 |

**Table 6.** Probabilities for risk scores in Experiment 3.

| Predicted level of threat | True threat status | | | |
|---|---|---|---|---|
| | Low ($V_1$) | Medium ($V_2$) | High ($V_3$) | |
| Score | (species $T_4$) | (species $T_3$) | (species $T_1$, $T_2$) | Row sum |
| 0  (very low) | 0.0710 | 0.0414 | 0.0414 | 0.1538 |
| 1  (low) | 0.0178 | 0.0533 | 0.0533 | 0.1243 |
| 2  (neutral) | 0.0473 | 0.0592 | 0.1538 | 0.2604 |
| 3  (high) | 0.0414 | 0.0355 | 0.1183 | 0.1953 |
| 4  (very high) | 0.0237 | 0.0651 | 0.1775 | 0.2663 |
| Column sum | 0.2012 | 0.2544 | 0.5444 | 1.0000 |

Before we make a prediction, all we know are the column sums: the relative frequencies of low ($V_1$), medium ($V_2$) and high ($V_3$) threat status among the species assessed. These are taken as the prior probabilities (Pr($V_1$) = 0.2012, Pr($V_2$) = 0.2544, and Pr($V_3$) = 0.5444). Suppose then that the predicted level of threat is *very high*. This prediction changes the prior probabilities to posterior probabilities Pr($V_1$|very high) = 0.0237/0.2263 = 0.0889, Pr($V_2$|very high) = 0.0651/0.2663 = 0.2444, and Pr($V_3$|very high) = 0.1775/0.2663 = 0.6667. We see that the effect of the prediction that the level of threat is very high for a species is to increase the probability that the species is *high threat*. The probability that the species is *medium threat* is little changed by the prediction that the level of threat is very high, while the probability that the species is *low threat* is reduced.

### 5.4. Experiment 1 revisited

Note that we can analyse Experiment 1 in the same way. From the frequencies given in section 3.1, we can calculate the corresponding probabilities (Table 7).

**Table 7.** Probabilities for risk scores in Experiment 1.

| Predicted level of threat | True threat status | | |
|---|---|---|---|
| Score $s_k(v)$ | Low ($\neg V$) | High ($V$) | Row sum |
| 0-3 ($k$=1) | 0.225 | 0.275 | 0.5 |
| 4 ($k$=2) | 0.150 | 0.350 | 0.5 |
| Column sum | 0.375 | 0.625 | 1.0 |

The prior probabilities are $\Pr(V) = 25/40 = 0.625$ and $\Pr(\neg V) = 15/40 = 0.375$. The posterior probabilities are then:

$\Pr(V|s_2(v)) = 0.35/0.5 = 0.70$
$\Pr(\neg V|s_2(v)) = 0.15/0.5 = 0.30$
$\Pr(\neg V|s_1(v)) = 0.225/0.5 = 0.45$
$\Pr(V|s_1(v)) = 0.275/0.5 = 0.55$

as in section 3.2.